\documentclass{an}
\usepackage{graphicx}
\usepackage{times}
\usepackage{cite}
\usepackage{multirow}
\newcommand{\ha}{H$\alpha$}

\begin{document}

\Pagespan{409}{416}
\Yearpublication{2014}%
\Yearsubmission{2013}%
\Month{12}%
\Volume{335}%
\Issue{4}%
\DOI{10.1002/asna.201312048}%

\title{Transmission profile of the Dutch Open Telescope H$\alpha$ Lyot filter}

\author{J. Koza\inst{1}\fnmsep\thanks{Corresponding author: koza@astro.sk}
  \and
  R.H. Hammerschlag\inst{2}
  \and
  J. Ryb\'{a}k\inst{1}
  \and
  P. G\"{o}m\"{o}ry\inst{1}
  \and
  A. Ku\v{c}era\inst{1}
  \and
  P. Schwartz\inst{1}
}

\titlerunning{Transmission profile of the DOT \ha\ Lyot filter}
\authorrunning{J. Koza et al.}
\institute{
  Astronomical Institute of the Slovak Academy of Sciences, 059 60
  Tatransk\'{a} Lomnica, The Slovak Republic
  \and
  Leiden Observatory, P.O.Box 9513, 2300 RA Leiden, The Netherlands
}

\received{2013 Dec 12}
\accepted{2014 Feb 24}
\publonline{2014 May 2}


\abstract{
Accurate knowledge of the spectral transmission profile of a Lyot
filter is important, in particular in comparing observations with
simulated data.
The paper summarizes available facts about the transmission profile of
the DOT \ha\ Lyot filter pointing to a discrepancy between
sidelobe-free Gaussian-like profile measured spectroscopically and
signatures of possible leakage of parasitic continuum light in DOT
\ha\ images.
We compute wing-to-center intensity ratios resulting from convolutions
of Gaussian and square of the sinc function with the \ha\ atlas profile
and compare them with the ratios derived from observations of the
quiet Sun chromosphere at disk center. We interpret discrepancies
between the anticipated and observed ratios and the sharp limb visible
in the DOT \ha\ image as an indication of possible leakage of
parasitic continuum light. A method suggested here can be applied also
to indirect testing of transmission profiles of other Lyot filters.
We suggest two theoretical transmission profiles of the DOT \ha\ Lyot
filter which should be considered as the best available
approximations.
Conclusive answer can only be given by spectroscopic re-measurement of
the filter.}

\maketitle

\section{Introduction}

Since the invention by Lyot (1933) and independently by \"{O}hman
(1938), the Lyot filter, called also the Lyot-\"{O}hman filter, the
birefringent filter, or less frequently the polarization-interference
monochromator (Stix 2004), earned broad utilization as an imaging
device in quasi-monochromatic wide-field surveys of the solar
atmosphere. An essential characteristic of any filter is its
transmission profile. Accurate knowledge of the profile is vital in
interpreting and comparing observations with simulated data. Methods
and results of measurement of transmission profile of Lyot filters are
given in van Griethuysen \& Houtgast (1959), Ramsay, Norton \&
Mugridge (1968), and Krafft (1968) showing also changes of the profile
with tuning of the filter. The general operating principles of Lyot
filters are introduced, e.g., in Title \& Rosenberg (1981) and
Bland-Hawthorn et al. (2001).

\begin{figure}
\centering
  \includegraphics[width=\columnwidth]{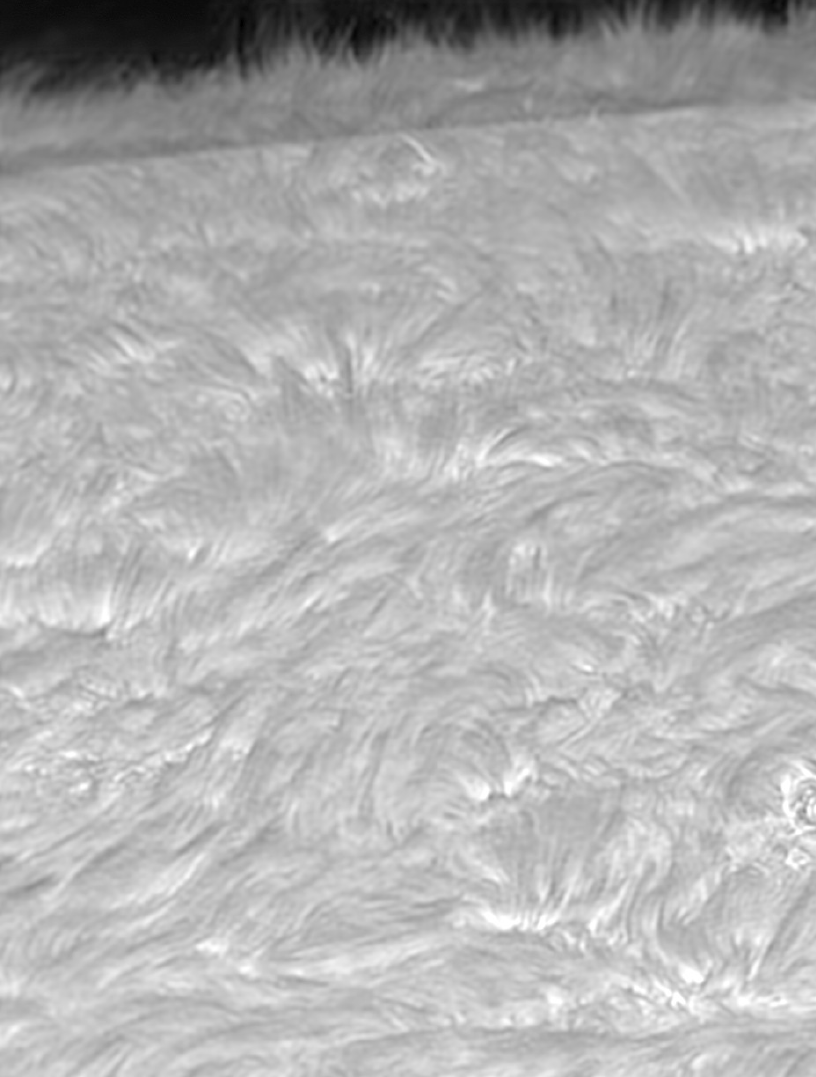} 
  \caption{DOT limb image taken on 2005 October 4 at 09:38:40\,UT
    in the \ha\ line center. Field of view: $66\times87$\,arcsec$^2$.}
  \label{fig1}
\end{figure}

The open database of the Dutch Open
Telescope\footnote{http://dotdb.strw.leidenuniv.nl/DOT/} (DOT;
Hammerschlag \& Bettonvil 1998; Bettonvil et al. 2003; Rutten et
al. 2004) offers many ready-to-use time sequences of
speckle-reconstructed \ha\ images obtained from 2004 to 2007 by a
tunable Lyot filter described in Gaizauskas (1976) and Bettonvil et
al. (2006). Fig.\,\ref{fig1} is an example of DOT image taken at the
limb in the \ha\ line center showing a thick hedge-row of spicules. It
is possible that the database will later be supplemented with
\ha\ time sequences obtained after 2007 since the DOT \ha\ data taken
in 2010 have already appeared in Rutten \& Uitenbroek (2012), Joshi et
al. (2013), and in the poster presentation by Aparna, Hardersen \&
Martin in the meeting of the Solar Physics Division of the American
Astronomical Society in 2013. Leenaarts et al. (2006) performed
spectral synthesis of the \ha\ line using a snapshot from 3D MHD
simulations and compared the results with DOT observations assuming a
Gaussian transmission profile of the DOT \ha\ Lyot filter.

An aim of this paper is to summarize available unpublished facts about
the transmission profile of the DOT \ha\ Lyot filter and confront them
quantitatively with observations and suspicions which appeared
recently in the literature. To reconcile an existing discrepancy we
suggest two theoretical transmission profiles of the DOT \ha\ Lyot
filter more compatible with observations. A method suggested here can
be applied to indirect testing of transmission profiles of Lyot
filters implemented, e.g., in the Narrow-band Filter Imager on-board
Hinode (Kosugi et al. 2007) or the Coronal Multichannel Polarimeter
installed recently on the Lomnicky Peak Observatory (Ku\v{c}era et
al. 2010; Schwartz et al. 2012).

\begin{figure}[!ht]
\centering
\includegraphics[bb = 20 0 170 201, width=\columnwidth, clip=]{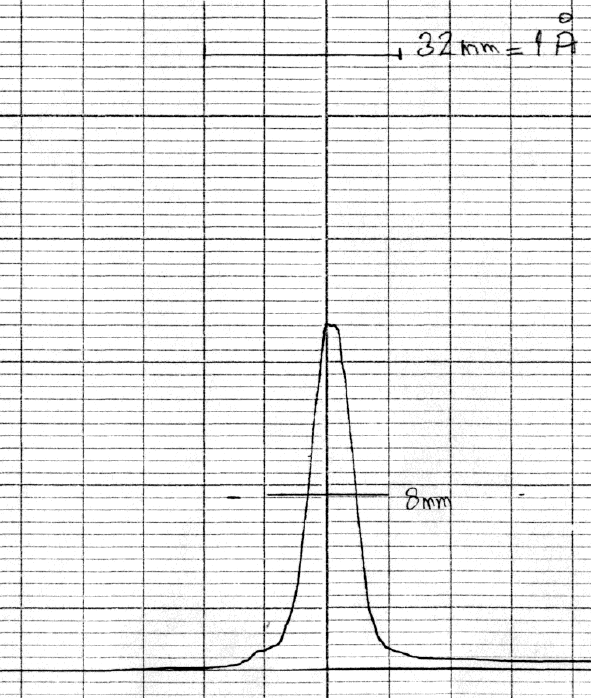}\\[1mm]
\includegraphics[width=\columnwidth]{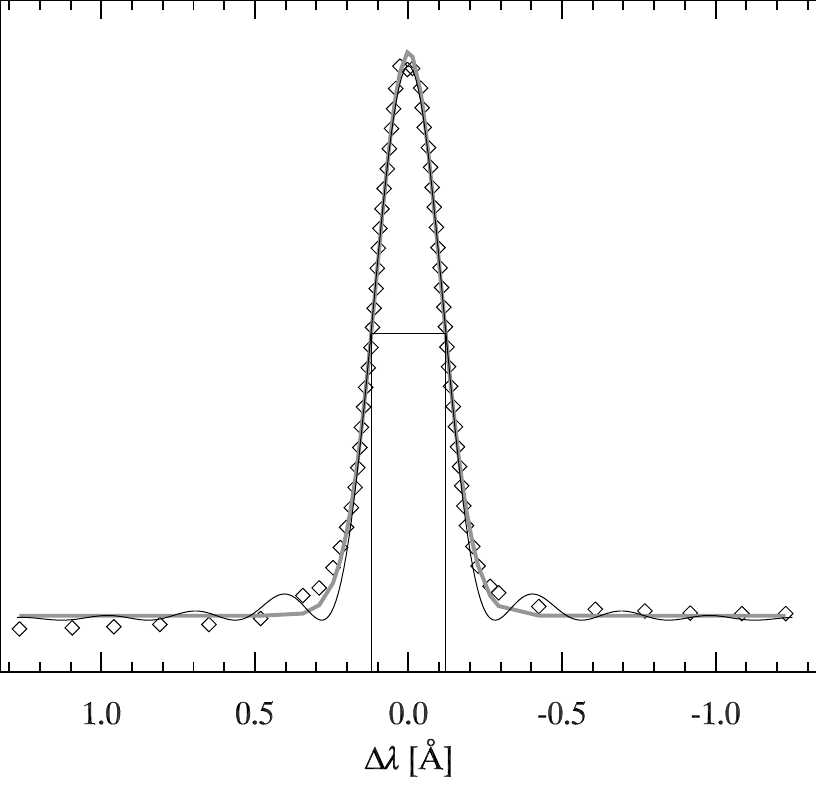}
  \caption{{\it Top}: scanned registration from ink recorder showing
    measured transmission profile of the DOT \ha\ Lyot
    filter. Distance of vertical lines corresponds to 10\,mm. The
    wavelength scale 32\,mm $= 1$\,\AA\ is handwritten in the upper
    right corner. Filter FWHM of 250\,m\AA\ is marked by the
    horizontal line followed by the flank-to-flank distance of
    8\,mm. {\it Bottom}: digitized transmission profile of the DOT
    \ha\ Lyot filter (diamonds), its Gaussian fit (thick gray line),
    and corresponding sinc$^2$ function (black line,
    Eq.\,(\ref{eq2})). Thin lines mark Gaussian FWHM of 243\,m\AA.}
  \label{fig2}
\vspace{-0.5cm}
\end{figure}

\begin{figure}
\centering
  \includegraphics[width=\columnwidth, clip = ]{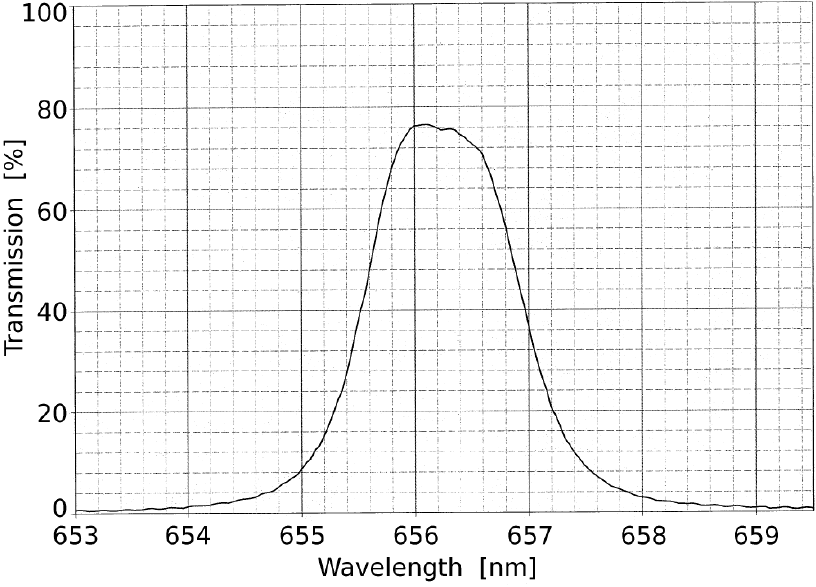} 
  \caption{Transmission of prefilter of the DOT \ha\ Lyot filter.}
  \label{fig3}
\end{figure}

\begin{figure}[!h]
\centering
  \includegraphics[width=0.99\columnwidth, clip = , angle=0.2]{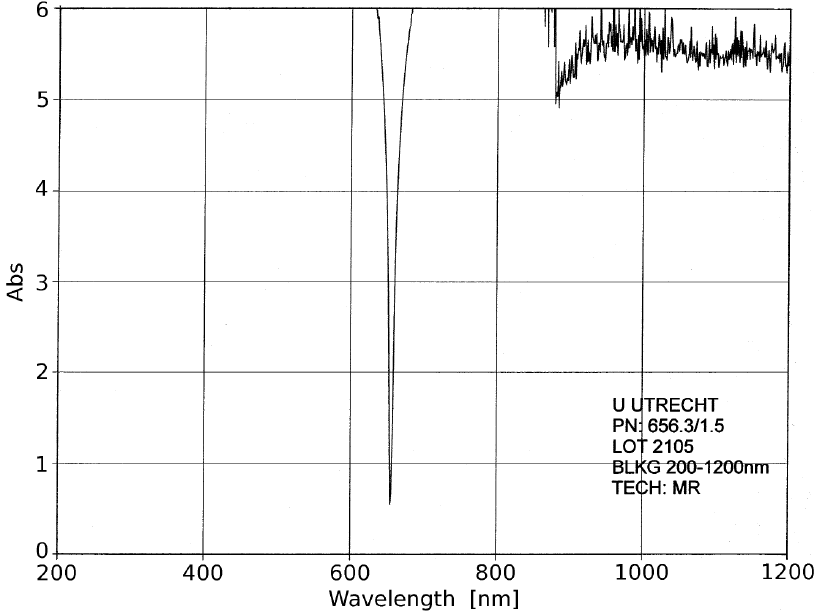} 
  \caption{Absolute value of the logarithm to base 10 of the prefilter
    transmission in Fig.\,\ref{fig3} over a broad spectral region. The
    transmission decreases upward.}
  \label{fig4}
\end{figure}

\begin{figure}
  \centering
  \includegraphics[width=\columnwidth,clip=]{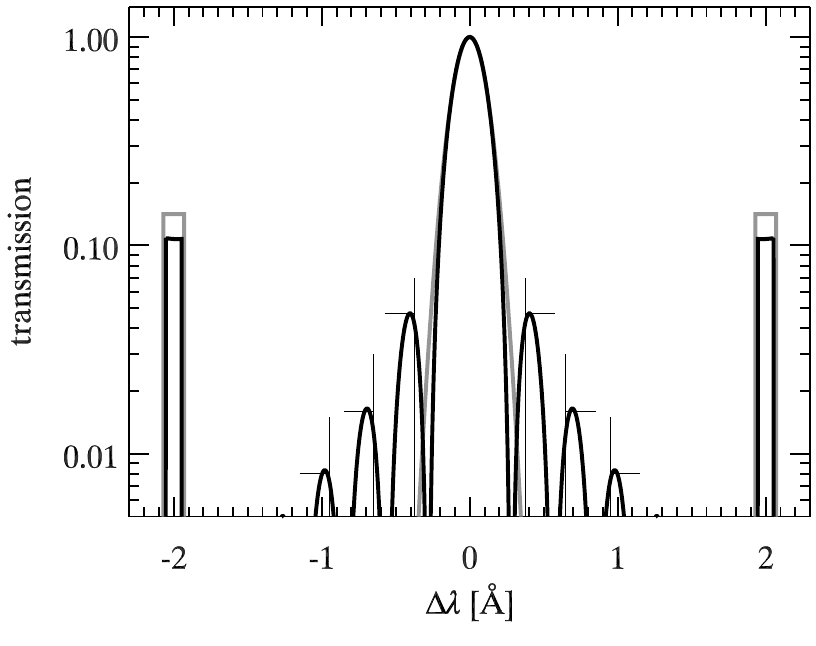} 
  \vspace{-0.8cm}
  \caption{The theoretical transmission profiles of the DOT \ha\ Lyot
    filter in the logarithmic scale: Gaussian $+ \Lambda$ (gray) and
    sinc$^2 + \Pi$ (black), both with FWHM = 250\,m\AA. The thin
    vertical lines with horizontal dashes approximately at $\pm 0.4,
    \pm 0.65$, and $\pm 0.95$\,\AA\ are the theoretical positions and
    the amplitudes of the subsidiary maxima according to Gaizauskas
    (1976, p. 8).}
  \label{fig5}
\end{figure}

\section{Spectroscopic investigation versus limb observation}

The convolved intensity $E(\lambda)$ of incident light with a
spectrum $I$ passing through a filter with a transmission profile $T$
centered at the wavelength $\lambda$ is the convolution
  \begin{equation}
    \label{eq1}
    E(\lambda) = \int_0^{\infty}I(x)T(x-\lambda)dx\,.
  \end{equation}
For an application of a Lyot filter as a spectroscopic device and
follow-up quantitative interpretation of observed $E(\lambda)$, e.g.,
through comparison with a synthetic spectrum $I$, one needs to assume
or to know the transmission profile $T$ and its possible variations
with tuning of the filter.

\renewcommand{\arraystretch}{1.2}
\tabcolsep=8.35pt
\begin{table*}
  \centering
  \caption{Column 1: ratios of convolved intensities
    $E(\Delta\lambda)$. Parenthesis $\langle \, \rangle$ indicate an
    average of intensities at $\Delta\lambda$. Columns 2--4: observed
    ratios of spatio-temporal means of DOT \ha\ datacubes obtained in
    the quiet Sun in the indicated days and average ratios of
    observations from 12 days. Columns 5--8: anticipated ratios
    computed by the atlas \ha\ profile and the particular transmission
    profile with FWHM = 250\,m\AA\ centered at $\Delta\lambda = 0, \pm
    0.35$, and $\pm 0.7$\,\AA\ from the center of the atlas
    profile. The symbols $\Lambda$ and $\Pi$ represent two rectangle
    add-ons of the Gaussian and sinc$^2$ function, respectively. See
    text and Fig.\,\ref{fig5}.}
  \label{tab1}
  \begin{tabular}{cccccccc}
    \hline
    \multirow{2}{*}{Ratio} & \multicolumn{3}{c}{DOT \ha\ Observations} & \multicolumn{4}{c}{Atlas \ha\ Profile + Transmission Profile:} \\
    &                2005 Oct 19                    &    2007 Sep 28      & 12-days average & Gauss & Gauss\,$ +\, \Lambda$    & sinc$^2$      & sinc$^2 + \Pi$   \\
    \hline
    $\langle E(\pm0.7) \rangle / E(0)$                                         &  2.32  & 2.34   & 2.34  & 3.28 & 2.35  &   2.78   &  2.35 \\
    $\langle E(\pm0.7) \rangle / \langle E(\pm0.35) \rangle$  &  1.75  & 1.75   & 1.75  & 2.10 & 1.77  &   1.94   & 1.76 \\
    $\langle E(\pm0.35) \rangle / E(0)$                                       &  1.33  & 1.34   & 1.34  & 1.56 & 1.33  &   1.43    &  1.34 \\
    \hline
  \end{tabular}
\end{table*}

The DOT \ha\ Lyot filter contains eight stages assembled from three
groups of birefringent quartz crystals and five groups of calcite
crystals (Bettonvil et al. 2006). Its transmission profile was
measured photometrically in 1999 by the solar spectrograph in
Sonnenborgh observatory in Utrecht (top panel of
Fig.\,\ref{fig2}). The measurement confirmed almost symmetric and
Gaussian-like transmission profile with the full width at half maximum
FWHM = 250\,m\AA\ without significant subsidiary maxima or far-center
sidelobes ruling out a leakage of unwanted parasitic light. It also
confirmed invariance of the profile in tuning. The bottom panel of
Fig.\,\ref{fig2} displays a digitized profile and its Gaussian fit
(thick gray line) with FWHM = 243\,m\AA. Note that the wavelength
increases from the right to the left and there is no transmission
scale since the measurement was not absolutely calibrated. In the
service deployment at DOT, the filter is preceded by a prefilter with
FWHM = 14.9\,\AA\ blocking out sidebands 128\,\AA\ apart from the
central transmission peak (Bettonvil et al. 2006). The latter value is
the free spectral range (FSR) of the filter. The transmission profile
of the prefilter delivered by its manufacturer is shown in
Figs.\,\ref{fig3} and \ref{fig4} for the narrow and broad spectral
range. Note in the latter figure the broadband transmission of
$10^{-5.5}$ in the IR spectral range spanning from 870 to 1200\,nm.

On the contrary, Rutten (2007, 2012, 2013) admits presence of
parasitic continuum light in DOT \ha\ images pointing to the double
limb (Rutten 2007) and the sharp limb (Rutten 2013) seen in the image
taken in the \ha\ line center (Fig.\,\ref{fig1}). In the figure the
limb shines clearly through the mass of spicules. Visibility of the
limb arc is highlighted by the dark band of variable width demarcating
its outer limit. The dark band and the limb visibility in the
\ha\ line center images taken by a Lyot filter have already been
described in Bray \& Loughhead (1974) identifying the latter as a
symptom of parasitic light. Suspicion of possible contamination of the
DOT \ha\ line center images by the parasitic continuum light may rise
if comparing Fig.\,\ref{fig1} with a limb image taken in the
\ha\ center by a Fabry-P\'{e}rot instrument. An example is the lower
right panel of Fig.\,6 in Puschmann et al. (2006) where the limb and
the dark band are much less apparent. Nevertheless, more similar
images taken with various Fabry-P\'{e}rot instruments at several
position angles along the limb would be needed for comparison.

\begin{figure}[!h]
  \centering
  \includegraphics[width=\columnwidth]{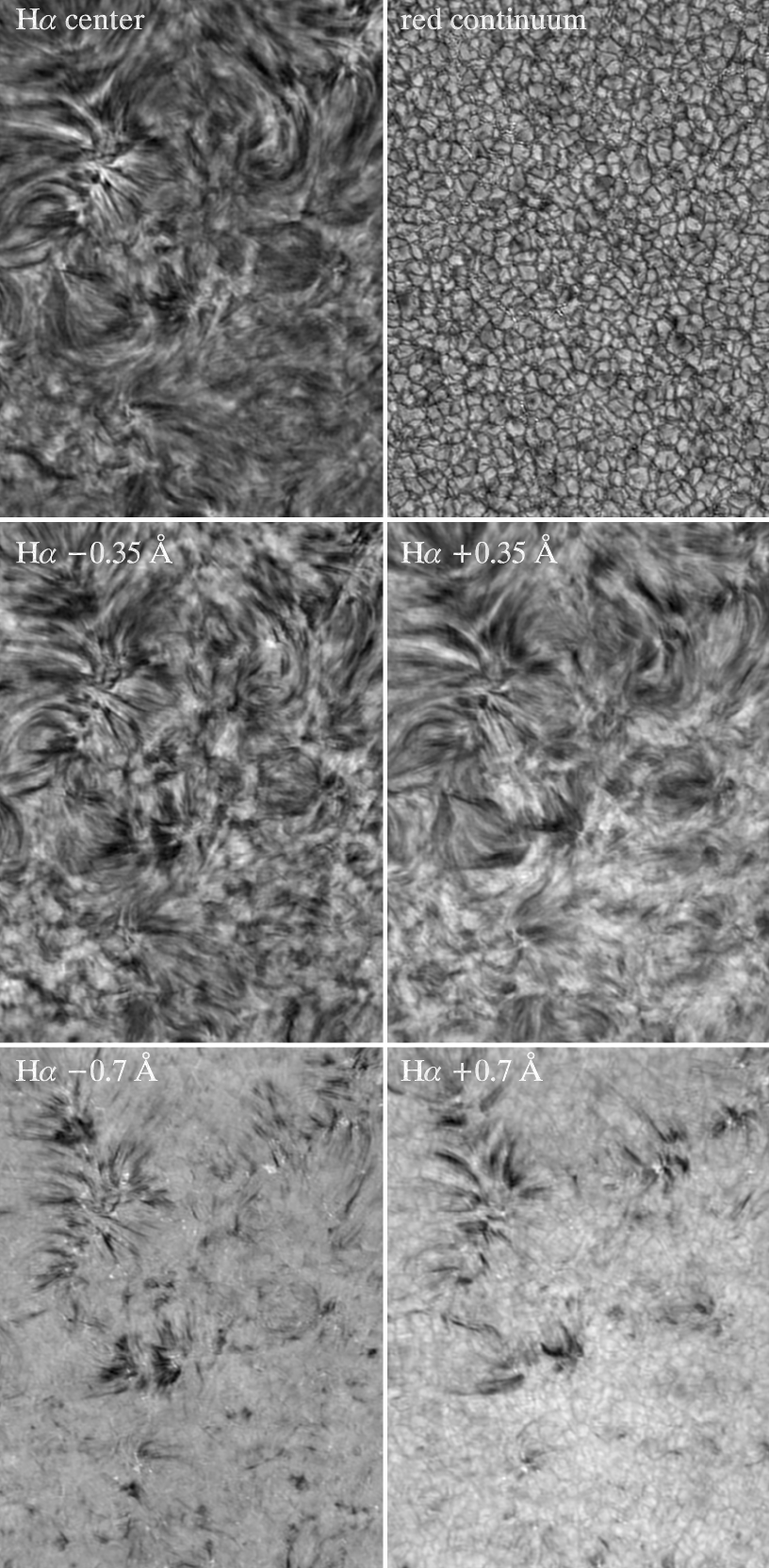} 
  \caption{Sample \ha\ images of the quiet Sun at the disk center
    recorded by DOT with the Hitachi KP-F100 camera with Sony ICX085
    CCD sensor (Rutten et al. 2004) on 2005 October 19 at 10:36:21\,UT
    at the moment of the best seeing occurred in the $41^{\rm st}$
    minute after the beginning of the observation at
    09:55:20\,UT. Field of view: $58\times79$\,arcsec$^2$.}
  \label{fig6}
\end{figure}

\begin{figure}[!h]
  \centering
  \includegraphics[width=\columnwidth]{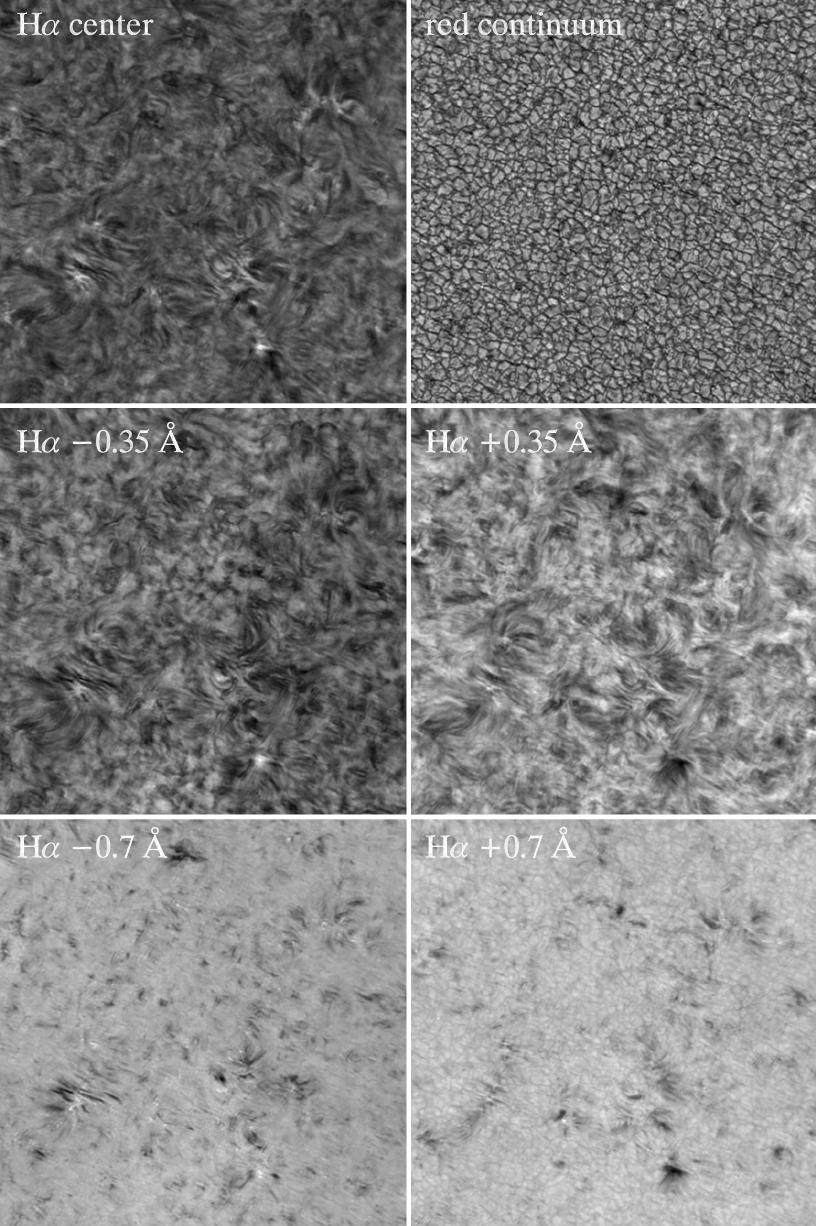} 
  \caption{Sample \ha\ images of the quiet Sun at the disk center
    recorded by DOT with the Redlake MegaPlus II ES4020 camera with
    Kodak KAI-4020 sensor on 2007 September 28 at 09:28:29\,UT at the
    moment of the best seeing occurred in the $53^{\rm rd}$ minute
    after the beginning of the observation at 08:35\,UT. Field of
    view: $93\times93$\,arcsec$^2$.}
  \label{fig7}
\end{figure}

\section{Towards new transmission profiles}

In this section we perform a quantitative indirect testing of the
transmission of the DOT \ha\ filter represented by the Gaussian with
FWHM = 250\,m\AA\ (Fig.\,\ref{fig2}) and a square of normalized sinc
function (from Latin {\it sinus cardinalis}, Fig.\,\ref{fig5}) as
suggested in Gaizauskas (1976). The latter is an approximation for
transmission of a single peak (see Appendix A) having the form
\begin{equation}
  \label{eq2}
        {\rm sinc}^2(\Delta\lambda) = \left( \frac{\sin\pi x}{\pi x} \right)^2,
\end{equation}
where $x = 2k\frac{\Delta\lambda}{\rm FWHM}, \Delta\lambda$ is a
distance from the center of the passband, and the transcendental
equation $(\sin\pi k)^2/(\pi k)^2=0.5$ yields the conversion factor $k
= 0.442946$ between $x, \Delta\lambda$, and FWHM/2. Since the function
has a singularity at $x=0$, its definition sets sinc$^2(0)=1$ for
$\Delta\lambda=0$.

First, we perform a simple check whether these theoretical
transmission profiles are compatible with observations. To this
purpose, we have chosen the \ha\ observations of very quiet areas at
the disk center available in the DOT database taken on 2005 October 19
and 2007 September 28 at the \ha\ line center and at $\pm 0.35$ and
$\pm 0.7$\,\AA\ off the center (Figs.\,\ref{fig6} and
\ref{fig7}). Quietness of the target areas is documented by the
continuum images taken simultaneously by a broadband interference
filter with FWHM = 2.4\,\AA\ centered at 6550.5\,\AA. The datasets
differ mainly in type of cameras used. We computed the spatio-temporal
mean of each datacube at the employed wavelength positions of the
filter and the ratios as defined in Col.~1 of
Table\,\ref{tab1}. Columns~2 and 3 show that these ratios are
independent of type of camera. They are also insensitive to the area
averaged since they do not change much if only large internetwork
areas are adopted for averaging. We checked also invariability of the
ratios with time taking single quiet-Sun \ha\ scans at the disk center
from twelve days mostly in 2007. Spans of the ratios are 2.25--2.44,
1.70--1.79, and 1.32--1.37 with averages given in Col.~4.

\begin{figure}
\centering
  \includegraphics[width=\columnwidth]{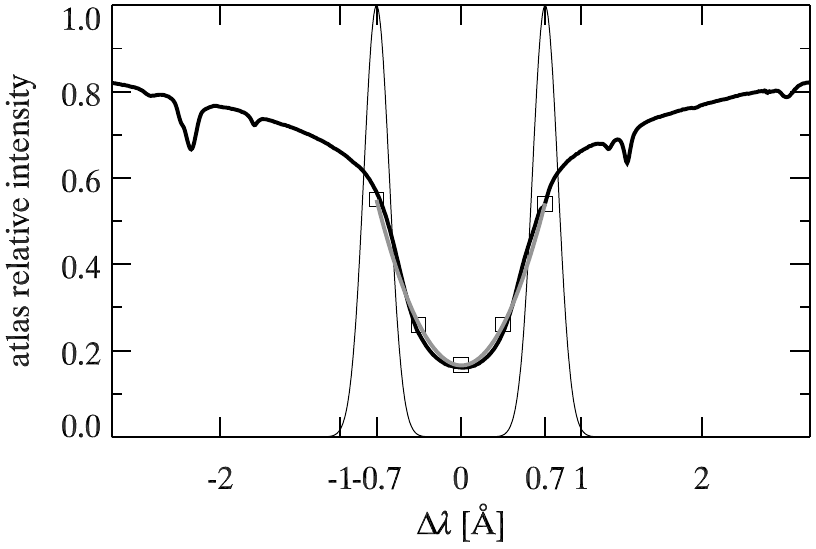} 
  \caption{Intensities (squares) obtained by convolution of the atlas
    \ha\ profile (thick black) and the Gaussian transmission profile
    with FWHM = 250\,m\AA\ centered at $0, \pm 0.35$, and $\pm
    0.7$\,\AA\ (the last two shown as thin solid) together with a
    parabolic fit (gray) of the intensities.}
  \label{fig8}
\end{figure}

\begin{figure}
\centering
  \includegraphics[width=\columnwidth]{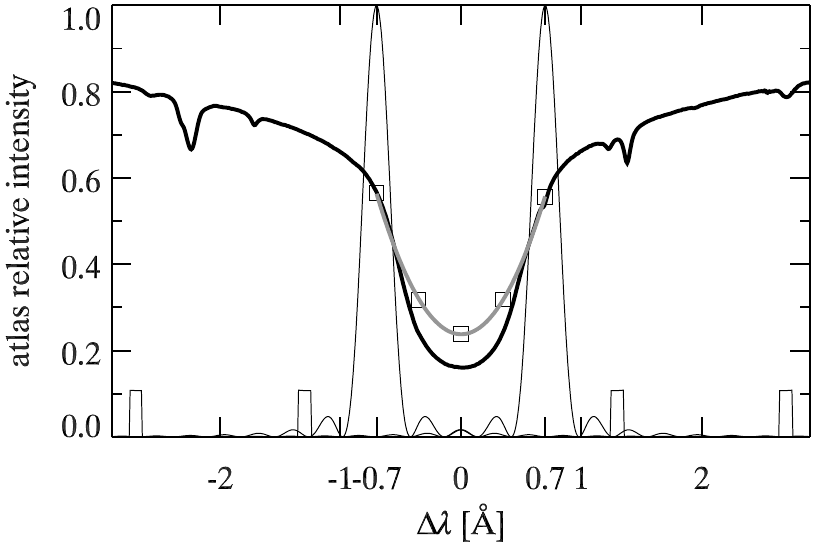} 
  \caption{The same as in Fig.\,\ref{fig8} but for the theoretical
    transmission profile sinc$^2 + \Pi$.}
  \label{fig9}
\end{figure}

Columns~5 and 7 of Table\,\ref{tab1} show anticipated ratios for the
\ha\ profile extracted from the spectral atlas (Neckel 1999) and
convolved with the Gaussian and sinc$^2$ function. Apparently, these
models of the transmission profile yield ratios significantly
exceeding the observed ones. It suggests in sharp contradiction with
the result of measurement in 1999 that the real transmission profile
of the DOT \ha\ filter might have larger throughput than these models
and lets in some parasitic continuum light contaminating mainly the
core and thus decreasing the observed ratios compared to the
anticipated ones. This is illustrated plainly in Figs.\,\ref{fig8} and
\ref{fig9} as the shallowing of the observed \ha\ core. Then the
additional continuum light in limb images taken in the \ha\ center
increases significantly the limb contrast with respect to off-limb
emission structures (Fig.\,\ref{fig1}) shining on the background of
scattered continuum light.

\renewcommand{\arraystretch}{1.2}
\tabcolsep=16.9pt
\begin{table}
\centering
\caption{Rectangle parameters.}
\label{tab2}
\begin{tabular}{cccc}
\hline
Rectangle &  Area   & Width   & Height \\ 
                & (m\AA) &  (m\AA) &          \\
\hline
$\Lambda$ & 20.0 & 141 & 0.141 \\
$\Pi$          & 11.5 & 107 & 0.107 \\
\hline
\end{tabular}
\end{table}

To account for the missing parasitic light, we constructed two models
combining the Gaussian and sinc$^2$ function (Eq.\,\ref{eq2}) with two
ad hoc rectangle functions $\Lambda$ and $\Pi$ (Fig.\,\ref{fig5})
centered at $\Delta\lambda=\pm 2$\,\AA\ around the \ha\ line
center. The symbols $\Lambda$ and $\Pi$ indicate the light leak and
the shape of the functions. The areas of rectangles were found by a
trial and error (Cols.~6 and 8 of Table\,\ref{tab1}) to match the
observed ratios. Parameters of a single rectangle are summarized in
Table\,\ref{tab2}. These extensions of Gaussian and Eq.\,(\ref{eq2}),
referred as Gauss\,$ +\, \Lambda$ and sinc$^2 + \Pi$, represent new
theoretical transmission profiles of the DOT \ha\ Lyot filter
accounting for the sharp limb in Fig.\,\ref{fig1} and reconciling
discrepancies between anticipated and observed intensity ratios in
Table\,\ref{tab1}. The integrals of the Gaussian and sinc$^2$
function, both with FWHM = 250\,m\AA, are 266 and 282\,m\AA\ (Appendix
B). For comparison, the rectangle add-ons increase their areas about
$40\times100/266=15\%$ and $23\times100/282 = 8\%$.

\section{Discussion}

Our study points out the striking discrepancy between the observed and
anticipated intensity ratios shown in Cols.~2--4 and Col.~5 of
Table\,\ref{tab1} assuming the Gaussian transmission profile in
concert with the measured Gaussian-like sidelobe-free transmission of
the DOT \ha\ Lyot filter (Fig.\,\ref{fig2}). Column~7 of
Table\,\ref{tab1} indicates some alleviation of the discrepancy for
the sinc$^2$ function (Fig.\,\ref{fig5}) suggested in Gaizauskas
(1976). Is it all truly an another manifestation of parasitic
continuum light visible in Fig.\,\ref{fig1} as the sharp limb as
suggested in Rutten (2007, 2012, 2013)?

On the suggestion of a referee, we discuss separately two likely leaks
of the parasitic light. These are the broadband transmission of
$10^{-5.5}$ of the prefilter in IR spanning from 870 to 1200\,nm
(Fig.\,\ref{fig4}) and/or the main passband of the Lyot filter
itself. Since original IR cut filters were dismantled from DOT cameras
prior their deployment and assuming that the DOT optics is fully
transparent in IR, the leak can be represented as an extension of the
right side of Eq.\,(\ref{eq1}) by $\varepsilon = \varepsilon_{\rm VIS}
+ \varepsilon_{\rm IR}$, where $\varepsilon$ is the total area of
rectangles in Table\,\ref{tab2}. Then $\varepsilon_{\rm VIS}$ and
$\varepsilon_{\rm IR}$ are the light leak through the main passband of
the filter and the IR leak of the prefilter. The latter can be
estimated by the formula
\begin{equation}
  \label{eq3}
        \varepsilon_{\rm IR} = \frac{F_{\rm IR}}{F_{\rm H\alpha}}\frac{P_{\rm IR}}{P_{\rm H\alpha}}\frac{S_{\rm IR}}{S_{\rm H\alpha}}\frac{\Delta\lambda_{\rm IR}}{FSR}\int_0^{\infty}T(\lambda)d\lambda\,,
\end{equation}
where $F_{\rm IR}$ and $F_{\rm H\alpha}$ are the average solar fluxes
in IR and the \ha\ continuum in the DOT altitude of 2350\,m for the
specific solar zenith angle. We computed their ratio of 0.45 by the
radiative transfer library
libRadtran\footnote{http://www.libradtran.org} (Mayer \& Kylling
2005). The factor $P_{\rm IR}/P_{\rm H\alpha}$ is the ratio of the
average transmissions of the prefilter in IR and \ha\ estimated from
Figs.\,\ref{fig3} and \ref{fig4} as $10^{-5.5}/0.76 \approx
4\times10^{-6}$. The next factor $S_{\rm IR}/S_{\rm H\alpha}$
accounts for different sensitivity of DOT cameras in IR and \ha. We
estimated its value of 0.12 from curves of spectral sensitivity of
their sensors showing that their efficiency has a cut off at
1000\,nm. The last but one factor $\Delta\lambda_{\rm IR}/FSR
\approx 10$ is the number of sidebands of the Lyot filter with $FSR =
12.8$\,nm and the integrated transmission of
$\int_0^{\infty}T(\lambda)d\lambda = 282$\,m\AA\ within the considered
spectral range of $\Delta\lambda_{\rm IR} = 1000 - 870 = 130$\,nm. The
product of these factors is $\varepsilon_{\rm IR} \approx
6.5\times10^{-4}$\,m\AA.

Consider now the extreme but unlikely situation that the polarizers in
the Lyot filter are completely ineffective in infrared wavelengths,
i.e., transmit all polarization directions equally, rendering the Lyot
filter ineffective. Then Eq.\,(\ref{eq3}) simplifies to the form
\begin{equation}
  \label{eq4}
        \varepsilon_{\rm IR} = \frac{F_{\rm IR}}{F_{\rm H\alpha}}\frac{P_{\rm IR}}{P_{\rm H\alpha}}\frac{S_{\rm IR}}{S_{\rm H\alpha}}\Delta\lambda_{\rm IR}\,,
\end{equation}
where the last factor $\Delta\lambda_{\rm IR} = 130$\,nm accounts for
complete ineffectiveness of the Lyot filter in the wavelength range
from 870 to 1000\,nm. For this extreme situation, we find that
$\varepsilon_{\rm IR}=0.3$\,m\AA\ which is only 1.3\% of the total
area of 23\,m\AA\ of two correcting $\Pi$ rectangles
(Table\,\ref{tab2}) proving that
\begin{itemize}
\item[--] the IR leak of the prefilter is negligible compared to $\varepsilon$
              listed in Table\,\ref{tab2} in the Area column,
\item[--] virtually all parasitic light leaks through the main passbands
              of the prefilter and the Lyot filter. 
\end{itemize}
For the latter, we can imagine the following possible source and cause
of the parasitic light
\begin{itemize}
\item[--] the central peak of the transmission profile is
  superimposed on the background of very low transmission spanning
  over the whole spectrum,
\item[--] the transmission profile has changed since its measurement in
  1999 and a new measurement would be needed.
\end{itemize}
The two rectangle add-ons of the Gaussian and sinc$^2$ functions
(Fig.\,\ref{fig5}) thus compensate for these effects or their
combination. A possible non-linearity of the cameras as a cause is
excluded since almost the same ratios are obtained for datasets taken
by different cameras (Figs.\,\ref{fig6} and \ref{fig7}, Cols.~2 and 3
in Table\,\ref{tab1}). Since the transmission measurement was not
absolutely calibrated (Fig.\,\ref{fig2}), the level of background
transmission is unknown. Nevertheless, a closer inspection of the top
panel of Fig.\,\ref{fig2} suggests an increased background
transmission extending blueward from the peak. This may contribute
significantly to the total amount of parasitic light.

Deviations in the relative rotational orientation of the eight
successive stages (Bettonvil et al. 2006) of the Lyot filter can be
the source of the blue-ward background. It did not change in value and
shape when the filter was tuned from $-1.5$ to $+1.5$\,\AA. This means
that these deviations occur in the initial rotational orientation of
the stages and do not change with the rotations for tuning, each next
stage with double speed of the preceding stage performed by
traditional toothed gear-wheels assuring no change in their relative
positions by tuning many times forward and backward.

The transmission curves of the 1999 measurements were scanned in the
spectrum using a moving slit with photomultiplier, which was sending
its signal to a recorder. The scanned curves were about 4\,\AA\ broad
with the transmission peak near the center. Consequently, the
indication of the blue-ward background is only visible till
2\,\AA\ from the transmission peak. The transmission region of the
prefilter is much broader, see Fig.\,\ref{fig3}. It is possible that
there is a red-side background further away from the transmission peak
of the Lyot filter. It can be caused by combinations of orientation
deviations of stages of higher and lower path differences between the
polarization directions. Consequently, the chosen correction with two
ad hoc rectangular functions at 2-\AA\ distance from the transmission
peak, well within the transmission range of the prefilter, is
rational.

Limb \ha\ images taken by a Fabry-P\'{e}rot instrument (FPI) offer a
possibility to simulate an influence of the parasitic light on
appearance of limb in DOT \ha\ line center images like
Fig.\,\ref{fig1}. The contamination can be simulated by summing of FPI
\ha\ line center image with an FPI continuum image but weighted
appropriately.  After some trial and errors one should arrive to a
small factor producing a sharp limb in the FPI \ha\ line center
image. We plan to perform this test in the future with images obtained
by some of high-performance Fabry-P\'{e}rot instruments.  The
most-probable source is the Interferometric Bidimensional Spectrometer
(Cavallini 2006) installed at the Dunn Solar Telescope. Its open
database\footnote{http://www4.nso.edu/staff/kreardon/dstservice/}
already offers potentially usable datasets taken in the service mode
in the first months of 2013.

\section{Conclusions}

In this paper we summarize available facts about the transmission
profile of the DOT \ha\ Lyot filter. Its accurate knowledge is
important, in particular in comparing observations with simulated data
as was carried out in Leenaarts et al. (2006). While the spectroscopic
measurement in 1999 showed almost symmetric and Gaussian-like
transmission profile without significant subsidiary maxima or
far-center sidelobes, two indirect and entirely different approaches
indicate possible leakage of parasitic continuum light into DOT
\ha\ images. To reconcile the discrepancy, we suggest two theoretical
transmission profiles of the DOT \ha\ Lyot filter combining the
Gaussian and sinc$^2$ functions (Eq.\,\ref{eq2}) with two ad hoc
rectangle functions. The extended Gauss$\,+\,\Lambda$ and sinc$^2 +
\Pi$ functions are equivalent. The former is simpler in use while the
latter conforms the theory (see Appendix A) but has a singularity at
$x=0$. These functions should be considered only as the best available
approximations usable in relevant applications. Decisive answer can
give only spectroscopic re-measurement of the transmission profile of
the DOT \ha\ Lyot filter yielding calibrated transmission.

\acknowledgements 
We are grateful to the referee for very helpful suggestions.  This
work was supported by the Slovak Research and Development Agency under
the contract No. APVV-0816-11. This work was supported by the Science
Grant Agency - project VEGA 2/0108/12. This article was created by the
realization of the project ITMS No.  26220120009, based on the
supporting operational Research and development program financed from
the European Regional Development Fund. The authors thank
P.~S\"{u}tterlin for the DOT observations and the data reduction. The
Technology Foundation STW in the Netherlands financially supported the
development and construction of the DOT and follow-up technical
developments. The DOT has been built by instrumentation groups of
Utrecht University and Delft University (DEMO) and several firms with
specialized tasks. The DOT is located at Observatorio del Roque de los
Muchachos (ORM) of Instituto de Astrof\'{\i}sica de Canarias
(IAC). DOT observations on 2005 October 19 have been funded by the
OPTICON Trans-national Access Programme and by the ESMN-European Solar
Magnetic Network - both programs of the EU FP6.

\appendix

\section{Mathematical background}
Lyot (1944, p.\,7) and Stix (2004, p.\,107) show that an incident wave
with the amplitude $A$ and the wavelength $\lambda$ exits from an
$N$-stage Lyot filter with the amplitude $A'$
\begin{eqnarray}
\label{eqa1}
A' & = & A \cos\frac{\pi e J}{\lambda} \cos\frac{2\pi e J}{\lambda} \cos\frac{4\pi e J}{\lambda}~\cdots~\cos\frac{2^{N-1}\pi e J}{\lambda} \nonumber \\
   & = & A \prod_{k=1}^N \cos\frac{2^{k-1} \pi e J}{\lambda}\,,
\end{eqnarray}
where $e$ is the thickness of the first crystal plate with the
birefringence $J$. Lyot (1944, p.\,7), van Griethuysen \& Houtgast
(1959, p.\,279), Title \& Rosenberg (1981, p.\,816), and
Bland-Hawthorn et al. (2001, p.\,615) express the exit amplitude $A'$
in the form
\begin{equation}
\label{eqa2}
A' = A\; \frac{ \sin 2^N \frac{\pi e J}{\lambda} }{2^N \sin \frac{\pi e J}{\lambda} }\,.
\end{equation}
We adopt in Eqs.\,(\ref{eqa1}) and (\ref{eqa2}) the same notation as
Stix (2004). Finally, Gaizauskas (1976, p.\,8) approximates the exit
amplitude $A'$ for a single peak with the formula
\begin{equation}
\label{eqa3}
A' = A\; \frac{ \sin \pi x }{ \pi x }\,.
\end{equation}

Since we did not find in available literature a derivation of
Eqs.\,(\ref{eqa2}) and (\ref{eqa3}), we show it here. Repeated use of
the double angle formula for the sine shows that
\begin{eqnarray}
\sin 2^1x     & = & 2^1 \sin x \cos x \,, \nonumber \\
\sin 2^2x & = & 2^2 \sin x \cos x \cos 2x \,, \nonumber \\
\sin 2^3x & = & 2^3 \sin x \cos x \cos 2x \cos 2^2x \,, \nonumber \\
\sin 2^4x & = & 2^4 \sin x \cos x \cos 2x \cos 2^2x \cos 2^3x \,, \nonumber \\
               & \vdots & \nonumber \\
\sin 2^Nx & = & 2^N \sin x \prod_{k=1}^N \cos 2^{k-1}x \,.
\end{eqnarray}
Then the product of the cosines is
\begin{equation}
\prod_{k=1}^N \cos 2^{k-1}x = \frac{ \sin 2^Nx }{ 2^N \sin x }\,,
\end{equation}
proving Eq.\,(\ref{eqa2}) for $x = \pi e J/\lambda$\,.

Let $E = 2^Ne = 2 \times 2^{N-1}e$ is a doubled thickness of the
$N$th crystal plate of a Lyot filter (Stix 2004). Exchanging $e$ 
for $E$ in Eq.\,(\ref{eqa2}) one can obtain
\begin{equation}
\label{eqa6}
A' = A\; \frac{ \sin \frac{\pi E J}{\lambda} }{2^N \sin \frac{\pi E J}{2^N \lambda} }\,.
\end{equation}
The limit of the right side of Eq.\,(\ref{eqa6}) for $N \to \infty$ is
\begin{equation}
\lim_{N \to \infty} \frac{ \sin \frac{\pi E J}{\lambda} }{2^N \sin \frac{\pi E J}{2^N \lambda} } =
\frac{ \sin \frac{\pi E J}{\lambda} }{\lim\limits_{N \to \infty} 2^N \sin \frac{\pi E J}{2^N \lambda} } =
\frac{ \sin \frac{\pi E J}{\lambda} }{ \frac{\pi E J}{\lambda} } \,,
\end{equation}
proving Eq.\,(\ref{eqa3}) for $x = \frac{EJ}{\lambda}$, because
$\lim\limits_{N \to \infty} 2^N \sin \frac{X}{2^N}=X$ (see, e.g.,
Morrison 1995).

Thus the single-peak approximation introduced in Gaizauskas (1976)
represents a hypothetical Lyot filter with infinite number of stages
with the full width at half maximum
\begin{equation}
{\rm FWHM} = 0.88 \frac{\lambda^2}{EJ}
\end{equation}
(see, e.g., Title \& Rosenberg 1981) but with the thickness of the
thinnest plate $e$ approaching to zero resulting in an
infinitely-large free spectral range defined in Title \& Rosenberg
(1981) as
\begin{equation}
{\rm FSR} = \frac{\lambda^2}{eJ}\,.
\end{equation}

\newpage

\section{Useful integrals}
\begin{equation}
\int_0^\infty e^{-(ax)^2} dx = \frac{\sqrt \pi}{2a}
\end{equation}

\begin{equation}
\int_0^\infty \frac{\sin^2(ax)}{x^2} dx = \frac{\pi}{2}|a|
\end{equation}

\end{document}